# 18

# Multi-Agent Systems and Blood Cell Formation

Bessonov Nikolai[1], Demin Ivan[2], Kurbatova Polina[2],
Pujo-Menjouet Laurent[2] and Volpert Vitaly[2]
[1]*Institute of Mechanical Engineering Problems, 199178 Saint Petersburg*
[2]*Université de Lyon, Université Lyon1, CNRS UMR 5208 Institut Camille Jordan*
*F - 69200 Villeurbanne Cedex,*
[1]*Russia*
[2]*France*

## 1. Introduction

The objective of this chapter is to give an insight of the mathematical modelling of hematopoiesis using multi-agent systems. Several questions may arise then: what is hematopoiesis and why is it interesting to study this problem from a mathematical point of view? Has the multi-agent system approach been the only attempt done until now? What does it bring more than other techniques? What were the results obtained? What is there left to do?

We hope that the following will give the reader all the answers to these questions. And even more, we would be delighted if after reading it, you would like to know more on this subject and try to work on it to contribute to the understanding of this complex field.

Let us start with the biological background in order to get a clear idea of the problem behind the model.

### 1.1 Hematopoiesis: what is it?

Hematopoiesis (from the ancient Greek meaning to make (ποιεῖν) blood (αἷμα)) is the scientific name used to describe the blood cell formation.

### 1.2 Where does it occur?

It appears in the yolk sac or blood islands during early embryogenesis. Then, with the development of the individual, it reaches the spleen, liver and lymph nodes to eventually settle down in the medulla, also known as bone marrow once this latter has been completely formed. This process takes place in the femur, tibia or any other long bones for children to finally moves to the pelvis, cranium, vertebrae and sternum in the adult bodies.

### 1.3 How does it work?

There are two main branches in hematopoiesis: myeloid and lymphoid (see Fig. 1). These two branches originate from the same cell type: the hematopoietic stem cells (HSC). The lymphoid branch gives birth to the T and B cells, antibodies and memory cells. Maturation, activation and some of proliferation of these latter are developed mostly into secondary lymphoid organs such as the spleen, thymus and lymph nodes. This is the reason why we shall not be focused on this branch. We might mention it from time to time though during this chapter when we would like to describe hematopoiesis in its whole.

Consequently, our main attention will be given to the myeloid branch. Three blood cell types arise from this branch through three cell lineages: red blood cells (erythrocytes), white blood cells and platelets (megakaryocytes) (see Fig. 1). Their daily production is fairly high: each second for instance, the body produces 2 millions of erythrocytes, also 2 millions of platelets and 700,000 granulocytes. Their lifetime differs from one type to another (120 days for erythrocytes, about 7 to 10 for the thrombocytes, and 6 to 14 hours only for the granulocytes (the shortest lifetime of these cell types).

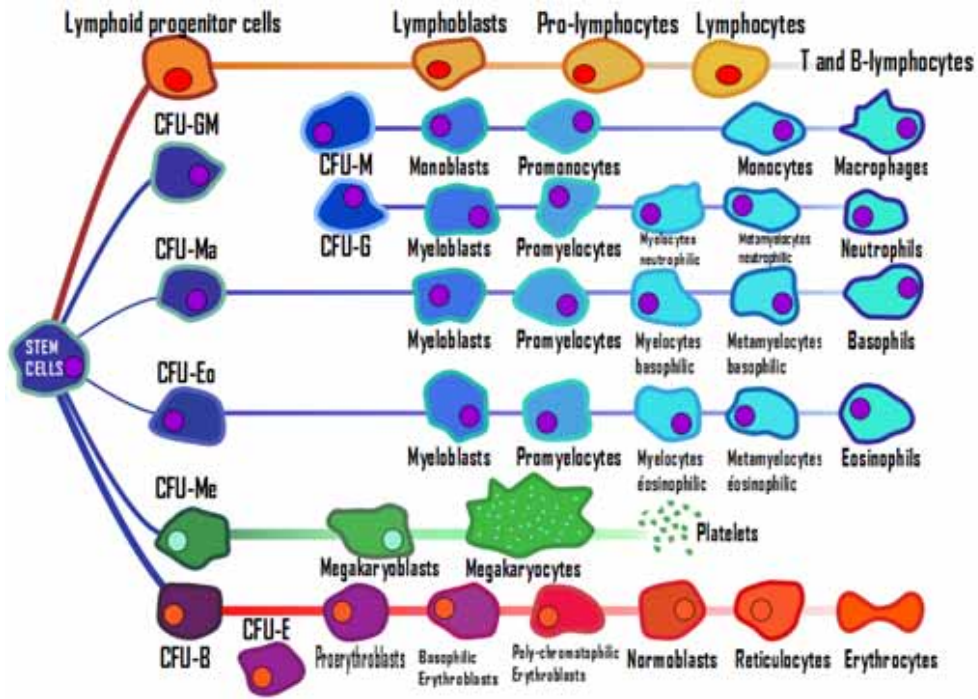

Fig. 1. Illustration of hematopoiesis: all blood cells originate from the stem cell compartment on the left and are released in the blood stream on the right. The lymphoid branch, on top, releases T and B-lymphocytes. The myeloid branch consists of the red lineage (bottom), white lineage in blue and platelets in green.

**1.4 The myeloid branch: an insight**

Let us have a closer look at the myeloid branch. But before doing this, it seems important to remind that all cells in each lineage originate from the HSC. These particular cells are able to self-renew. Their lifetime and number are still unknown, even if some attempts were done to predict their number in the body (Dingli et al., 2007a, b). Besides self-renewing, each stem cell can differentiate to a more mature cell, also called progenitor cells or it can die by apoptosis (natural cell death). HSC when differentiating give birth to early progenitor cells, too immature to belong to the myeloid or lymphoid branch. It might indeed be possible for them to move from one lineage to another one with no major difficulty. It is also possible for progenitor cells to self-renew under specific circumstances such as anemia, blood transfusion or any case of strong stress related to a loss of blood (see section 3.1.3). In normal cases progenitor cells are more incline to differentiate or die. It is only after few divisions that a cell reaches one of the three specific lineages and should not change its fate. These lineages are:

-the red blood cell lineage: progenitors are called CFU-B and CFU-E (where CFU stands for Colony Forming Units), their maturation evolves through different stages of precursor cells called erythroblast to finally become reticulocytes and eventually reach the blood stream under the form of erythrocytes, ready to transport oxygen.

-the megakaryocytic lineage: progenitors are called CFU-Me, and after having differentiated into megakaryoblasts, they become mature megakaryocytes. These cell types are really large (about 40 to 100 µm when the other blood cell sizes range between 1.5 to 24 µm). The megakaryocytes then split into hundreds of parts and give birth to platelets ready to reach the blood stream.

It is interesting to note that in the early stage of development, the young erythrocytes and megakaryocytes have a common root, the bipotential primitive megakaryocyte-erythroid precursor (MEP), located right after the stem cell compartment and right before the CFU-B and CFU-Me branches.

-the white blood cell lineage: the progenitors are located into three subgroups, CFU-GM (that gives also two other subgroups the CFU-M and CFU-G), CFU-Ma and CFU-Eo, who, after several differentiations and having proceeded through different stages (the "blast" precursors ones), respectively give birth to four white cell types: the macrophages, neutrophils, basophils and eosinophils, all of them ready to protect our body.

**1.5 How do these processes regulate?**

This is one of the key and challenging questions of this chapter. It is well known now that each lineage has at least one hormone regulating each lineage production. Indeed, several regulation factors are involved in the blood system to keep it in homeostasis. These controls are quite complex and many molecules and kinetic cascades are required. In this paragraph we describe only the main stimulating factor corresponding to each lineage. We refer the reader to more biologically detailed publications to get a better idea of all the chemical reactions and feedbacks implicated.

For the red blood cell lineage, this stimulating hormone is called erythropoietin (EPO). When produced in high quantity by the kidneys, it prevents the erythrocyte population from being lost by apoptosis. A large quantity of progenitor and precursor cells can then differentiate and a large quantity of erythrocytes would rapidly reach the blood stream. This happens in cases of important blood loss, anemia or any other stress erythropoiesis.

For platelets, the regulating factor is called thrombopoietin (TPO) and seems to target the differentiation of megakaryocytic. It is produced in the liver.

For the white cell population, the hormone is called granulocyte colony-stimulating factor (G-CSF). This molecule seems to stimulate survival, proliferation and maturation of the white cell progenitors and precursors. It is mainly produced by endothelium and macrophages and is overproduced in case of pathologies like neutropenia. See Fig. 2 for an overview of these different stimulating factors in the myeloid branch.

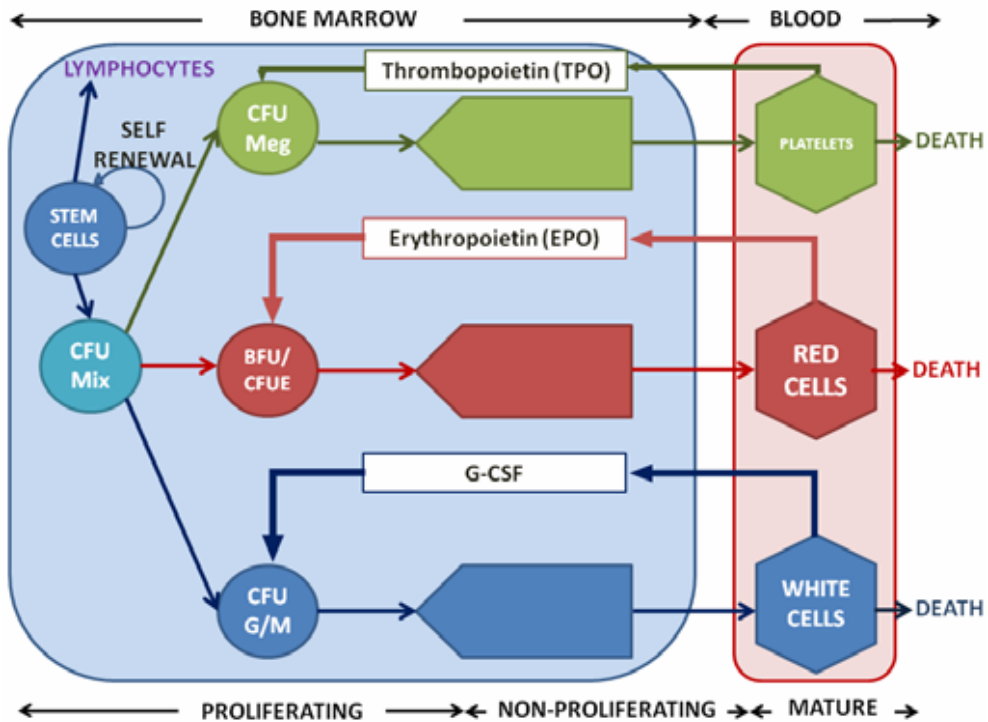

Fig. 2. Illustration of the differentiation pathways in the myeloid giving a representation of control of the platelet (in green), red blood cell (in red) and granulocyte or monocyte (G/M) (including neutrophils, basophils and eosinophils) production (in blue). The control loops, respectively mediated by thrombopoietin (TPO), erythropoietin (EPO) and the granulocyte colony stimulating factors (G-CSF) are indicated.

## 2. Objective of the chapter: what is the problem about?

Even if the regulating hormones presented in the previous paragraph are widely investigated in the biologist community, several questions and important issues remain open: how, for instance does thrombopoietin exactly act on the megakaryocytic lineage? Does it act on the apoptosis rate also, like EPO on erythrocytes or does it act only on the differentiation rate? What about G-CSF? Moreover, is spatial distribution of the cell in the bone marrow important or not in homeostasis? Do cells communicate between each other? If yes, how do they proceed? How do some diseases spread in the blood system, while some others do not? Is it possible for anyone to develop leukemia without knowing it, and to recover without any cure? How do stem cells and progenitor cells choose their lineage? Is this due to the environment of the cell, that is some external information or does it come from some stochasticity, some random noise inside the cell itself that leads its decision to prefer one lineage rather than another one?

Some of these questions have been tackled for almost 50 years now from a mathematical point of view, with different models and techniques. Some researchers used non linear partial differential equations with delay or not, some studied several systems of ordinary differential equations, linear or not, stochastic differential equations, etc. However, among all the existing models, and to the best of our knowledge, very few attempts were made to take the bone marrow structure and cell interaction into account.

The objective of this chapter is to try to answer some of the previous questions by the use of multi-agent systems and taking the bone marrow structure and space competition between cells into consideration. Before this, it seems necessary to justify the reason why the approach of the problem with the use of multi-agent system would be a good technique in comparison with other models. This is the reason why, in the next section we set up the state of the art, reminding some of the previous models and results obtained in the past decades. Then, we introduce different multi-agent approaches used to model hematopoiesis. In section 4, we give some of our contributions to this field and eventually conclude with what we believe was successful, what needs to be improved and the work planned for future investigations.

## 3. State of the art: different approach to model hematopoiesis

### 3.1 Deterministic models:
#### 3.1.1 The first models

Deterministic models are considered to be the first models describing cell cycle. In 1959, Lajtah et al. were the first to introduce a cell cycle model with a resting phase. Then, in the early 1970's, Burns and Tannock (1970) as well as Smith and Martin (1973) carried on Lajtah's work using a two phase model: the proliferating phase and the resting phase. This study was then generalized by Mackey in the late 1970's and applied to the study of hematopoietic stem cell. All these models consist in systems of ordinary differential equations, linear or not. In 1976, Wazewska-Czyzewska and Lasota (1976) proposed a similar model but they introduced a discrete model and applied it to erythrocyte production.

#### 3.1.2 Development of the models: going to more realistic and complex models

Each model afterward was more or less built from the first ones, with significant improvement, adding nonlinearity, delay. Several systems of partial differential equations arose then from that time; some were structured in age, size or maturity, sometimes two of them at the same time, sometimes with discrete delays, other times with distributed delays. They have been used to describe different parts of hematopoiesis: it could have been the stem cell compartment only, the red blood cell lineage, the myeloid lineage, with feedback or not regulating the production. The objective of each model was to understand the possible dysfunction of the system leading to diseases like anemia, leukemia, neutropenia or thrombocytopenia. Some of these diseases are chronic (this will be developed below) and oscillations of the size of cell populations could occur: this is the case for chronic myelogenous leukemia, cyclic neutropenia or cyclic thrombocytopenia. Incorporating one or several feedback loops on one or more lineages in the model was then necessary to simulate a possible regulation of the blood cell in the bone marrow.

#### 3.1.3 Some success stories

All these deterministic models combined with the study of the influence of different parameters allowed the authors of these researches to obtain accurate results and even predictions in rather good agreement with the experiments.

It has been proven for instance by Crauste et al. (2008) that EPO was not the only growth factor involved in the recovery of red blood cell homeostasis in case of stress erythropoiesis (anemia, important blood loss, blood transfusion, etc.). It was indeed shown mathematically that some glucocorticoids were necessary to explain the rapid recovery of normal red blood cell levels in mice suffering from severe loss of erythrocytes. It is indeed believed that glucocorticoids encourage progenitor cells to self-renew which does not happen or rarely in non pathological cases. In the same paper, the authors predicted also the fact that after such a pathological event, new born cells, too rapidly put in the blood stream were too weak to get a longer lifetime than usual. Consequently, the death rate of such cells was shown to be higher than the average normal erythrocytes.

Another prediction was made in a cyclic neutropenia model (Colijn et al., 2005a, b). The authors forecast that the G-CSF treatment (stimulating factor for the white cell lineage) would decrease the rate of apoptosis among neutrophil precursors back to normal levels while the differentiation rates for the neutrophil lineage would rise. It has also been assumed that apoptosis rate of the proliferating stem cells should be amplified for treated cyclical neutropenia (CN). Foley et al. in 2006 and Colijn et al. in 2005 (a, b) suggested then that it would be possible to get the same therapy effect with less G-CSF as usually used by changing the timing and duration of the treatment. This was obtained by combining the existing models on hematopoiesis with a model of G-CSF pharmacokinetics and changing the time interval between treatments and taking the time in the cycle into account.

Other approaches allowed nice discoveries: it was, for instance, possible to explain how such a short cycle of about 24 to 48 hours for blood cells could induce the oscillations of about 40 up to 80 days in the whole blood system in case of chronic myelogenous leukemia (CML). This work has been investigated by Pujo-Menjouet et al. (2005) and Adimy et al. (2007). The authors used different deterministic models and found two groups of parameters able to change either the period of the oscillations or their amplitude. These two groups were:

- the parameters involved in cell loss (apoptosis and differentiation) able to change the periods of the population dynamics, and
- the parameters involved in the cell cycle regulation (duration of the proliferating phase, and reentry rate of the resting cells into proliferation) able to change the amplitude of the cell population.

Many other results were achieved with deterministic models and one of the latest models try to include treatment strategies using not only the stimulating factors such as EPO or G-CSF as mentioned above but also some drug associated to chemotherapy such as Imatinib to treat certain types of leukemia (Michor et al. (2007a, b)).

However, it seems important to note that something quite relevant is missing in all the models cited in this section. Not a single model here takes the bone structure into account. Moreover, there are three big issues that should be incorporated in the studies: how to cell communicate between each other and get the information from the environment? How do they decide which lineage should be chosen when they are still immature enough to decide and be able to change from one branch to another one (like the MEP presented in section 1.4)? Finally, what could be said about the niches (places seeded by stem cells to give birth to different colonies of all blood cell types)? Is the number of cells in this niche large enough to consider continuous deterministic models, or is it possible to describe these niches with other discrete and stochastic models taking space competition into account?

The last question would be answered yes. Under the assumption that a study is mainly focused on very few niches at the same time in a tiny part of the medulla; a discrete and stochastic approach would make sense.
This is what is developed in the next paragraph.

### 3.2 A different approach: the multi-agent models
Before the introduction of the multi-agent models, it appears necessary to have an insight of what has been done in term of stochastic models. They have been used to describe the mechanism of cell proliferation determined with a certain probability, and not by previous deterministic mechanisms. It is also important to remind one of the rare existing models with reaction diffusion taking the spatial structure of the bone marrow into account. This will help to justify the use of our software based on the multi-agent systems taking the medulla structure into account.

#### 3.2.1 Stochastic models
In every deterministic model, any cell fate such as differentiation, apoptosis or self-renewal is predicted by specific processes well defined, like a good engine that self-regulates. In case of deregulation, the whole mechanism reacts and tries to reach back its non pathological equilibrium, also called homeostasis. Sometimes things do not occur in this way, and other equilibria can be reached, changing the population fate and subsequently the whole system.
However, in vivo, the cell decisions may not originate from well determined laws, and the parameters involved in the problem can exhibit great sensitivities to tiny changes. These small variations could appear in the inside of the cell (intrinsic) as well as its external environment (extrinsic). This problem has been investigated in the study of stem cells in the late 1990's and year 2000's with the work of Abkowitz (1996, 2000), Dingli, (2006, 2007a, b), Newton (1995), Roeder and Loeffler (2002, 2006b). The authors used discrete models where decisions ruling the cell future could be made following stochastic processes.
Some studies have shown the high sensitivity of stability of the HSC system to perturbation and death rates but not to proliferation rates (Lei and Mackey, 2007). The influence of extrinsic fluctuation has been modelled by Gillespie (1992) and Shahrezael (2008). Concerning the intrinsic perturbation, the influence of intern information and variation inside the cell nucleus leading to a drastic change of its fate is still in discussion amongst biologists. It is currently being investigated by mathematicians who would like to understand the influence of these changes to the lineage choice of a progenitor cells due to the different changes occurring randomly in the nucleus (variation during transcription of a gene, translation or mRNA, etc.). Sensitivity to such modifications would decrease as cells increase their maturity. In other words, it would be more difficult for a precursor cell to change its lineage, while, an immature cell, let say a MEP progenitor could be easily influenced to become either a megakaryocyte or an erythrocyte. This decision could occur as explained above at the molecular level, when stochasticity would have a greater influence. Thus multi-scale models would be necessary. This has been already proposed by Crauste et al. in 2010.
All these works are of great interest, but still, one important thing is missing in the models: space. Consequently, cell competition for space, their communication depending on their position, and of course the bone marrow structure should be taken into account. However, one deterministic approach exists and is briefly explained in the next paragraph.

### 3.2.2 Spatial models

In 2005, Bessonov et al. proposed a spatial model in order to describe the influence of the medullar stroma. Indeed, in these compact areas, spatial position and competition are important. This is even more a crucial matter in case for instance of acute leukemia. When this pathology develops in the bone marrow, immature cells, mostly white cells, overtake the space dedicated to more mature cells. These latter are pushed out from the marrow directly to the blood stream without completing their maturation process. The whole system is rapidly invaded by immature cells unable to satisfy the function requested. Furthermore, they stop the development of cells from other lineages which causes anemia due to a lack of erythrocytes and hemorrhages because of a lack of platelets. The approach introduced by Bessonov et al. in 2005 consists in a simplified continuous model describing cell movement in the stroma. It is built with reaction-diffusion systems of partial differential equations with convection. The role of cell diffusion was used to illustrate a random motion in the stroma, mechanical pressure between cells was set up explain the space competition in the marrow and Darcy law in porous medium allowed the authors to simulate the medullar stroma. Existence of a diffusion threshold for leukemic cells was proven, below which the healthy state loses its stability and let the leukemic cells overtake the system. Their simulations showed also the action of chemotherapy on the proliferation velocity of the cells.

### 3.2.3 Multi-agent models: a compromise

There exist two ways to combine spatial models with stochasticity. One way could be to include some stochasticity in the continuous reaction-diffusion system of partial differential equations. The second way would be to consider discrete multi-agent models. To the best of our knowledge, the first way has not been tackled yet. This is the reason why we focus our attention here on the second approach: the multi-agent models.

The main objective of the use of the multi-agent systems applied to hematopoiesis is to simulate cells as individual capable of self-renewal, apoptosis or differentiation in a closed space representing a part of the bone marrow. The first models appeared in 2006 with Pimentel 2006 who introduced a simple interface based on the early 1978's Mackey model on hematopoiesis. In 2008, D'Inverno et al. worked on a multi-agent model simulating stem cells but the problem was more adapted for the intestinal crypt cells.

Ramas at the same period developed a software package named Netlogo (http://ccl.northwestern.edu/netlogo/), a "programmable modeling environment for simulating natural and social phenomena", with one application to the blood cell formation. However, Netlogo's aim is not to model hematopoiesis only. Thus, many specificities related to the bone marrow do not appear. This is the reason why in 2006, Bessonov et al. created a new multi-agent based software dedicated only to the cell interaction in the bone marrow. This work integrated complex processes that have not been taken into account by the previous studies, such as cell communication, size difference, cell differentiation, space competition, pathological and non pathological cell mutations, spread of diseases like leukemia and the bone marrow niche.

All the details of this new interface are developed in the next section.

## 4. A specific multi-agent model adapted to hematopoiesis

How is it possible to model hematopoiesis in the bone marrow in a realistic way using at the same time the space structure and the cell population dynamics? The aim of this section is to present an attempt to answer this question by the use of the multi-agent systems. It has been introduced by Bessonov et al. for the first time in 2006. An improved version of the software was given in Bessonov et al. in 2009 and new perspective applications are introduced at the end of this section.

### 4.1 The software basis
Before showing simulations of the software we propose the basic assumptions made to obtain a clear, realistic and understandable approach of the problem. It appears necessary thus for a start to expose the way the cell cycle was modeled. Then it seems interesting to get an insight of how the bone marrow structure could be described and how the space competition could be simulated.

### 4.1.1 Modeling the cell cycle
In the software it was necessary to depict precisely the different cell fates. We assume here that cells are small disks having three possibilities: self renew (a capacity mostly authorized for stem cells, but this can be applied to any other cell types, such as progenitors, precursors or mutated cells), differentiate or die by apoptosis (natural death).

4.1.1.1 The division process

Before dividing, a cell will grow to get enough room for its two daughter cells. Note that we assume here that a mother cell will give birth to two daughter cells. But this is not exclusive. It is possible in the software to let a cell be able to divide into more than two cells: this could be used to get a faster result in the simulations. A stem cell could for instance give birth to 4 cells, one would be another stem cell, and the other three cells would correspond to the progenitors of the three lineages of the myeloid branch (see Fig. 3). Each cell can be given a specific color depending on its type: yellow for instance for stem cells, red for the red lineage, green for platelets and blue for the white lineage. This depends totally on the user's decision. Each lineage can be simulated at the same time or not. One can focus only on one branch, each one having a specific property or not: the size of the white cells or erythrocytes decreases as their maturity increases but this is exactly the opposite for megakaryocytes for instance. The size of each cell type can be determined by the user just by changing the radius of the disks illustrating the cells.

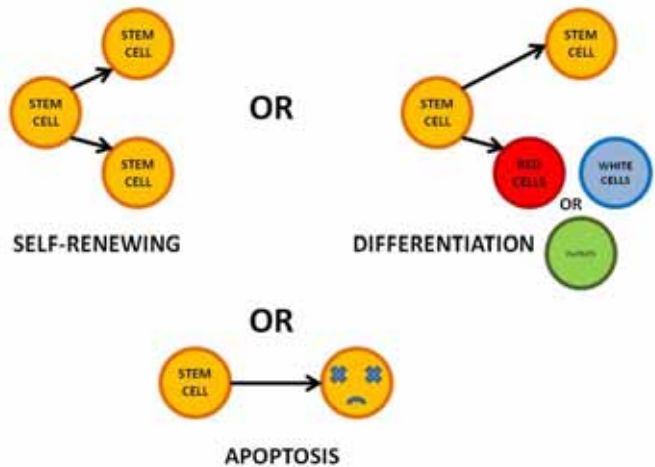

Fig. 3. Different stem cell fates taken into account by the software. A stem cell in yellow can either give birth to two daughter cells with exactly the same structure as their mother, or give birth to one stem cell and one differentiated progenitor (red cell, white cell or platelet), or die by apoptosis.

4.1.1.2 Stochasticity in proliferation

A cell can die by apoptosis at any time. The proliferation duration can be constant and defined by the user for each cell type, but it can be determined and occur within a time distribution: a constant time plus or minus a range whose value is defined by the user. This is one of the various specificities of this software. It is also possible to decide what could be the probability for a stem cell for instance to give birth to a red white or platelet progenitor. This ability of the software seems quite important for hematologists in the sense that almost 45 % of the blood consists of red blood cells that is about 99% of the hematocrit (portion of cells in the blood, the other portion consist of plasma, that is the remaining 55% of blood). The rest of the hematocrit is composed of white cells for 0.2%, and megakaryocytes between 0.6 to 1%. Thus, it seems realistic to assume that a stem cell has more chances to give birth to a red blood cell rather than a cell from another lineage. The software offers this possibility by choosing different probability for a cell to give birth to a certain cell type. This possibility includes also the probability to die. Thus, even apoptosis can be given a random rate that can be determined by the user (see Fig. 4). This is also relevant in the sense that apoptosis is rather important in the erythrocyte lineage, and this rate can be reduced under the influence of EPO. The influence of EPO and other simulating factors will be described later on.

### 4.1.2 The bone marrow structure

The bone marrow is set up as a simple rectangle in the software. Any time that a cell is pushed away from the rectangle border, it is assumed to reach the blood stream. The size of the rectangle can be chosen easily by the users, and modified anytime. Moreover, in order to be more realistic, it is possible to introduce fixed segments of different size anywhere in the rectangle to simulate the porosity of the bone marrow. These segments cannot be crossed by the cells and they are considered as walls necessary to bypass for the cells. The user can place the segments in different ways: they can be put like a bottle neck to force the cells to use only one way out to the blood stream, they can be in 3 of the 4 borders of the rectangle to give only one possible side for cells to reach the stream, they can also represent different niches where stem cells could develop colonies forming grapes of new born cells. Viscosity of the blood cells in the bone marrow can also be decided by changing a parameter value in the run window of the software.

When dividing, each cell giving birth to two or more daughter cells pushes away the other individuals. Space competition is then described. Consequently, if one cell type divides more rapidly than others, the bone marrow would swiftly be full of this type of cells and offspring, the other cells would be pushed away out in the bone marrow, or would have no room to develop their lineage. This phenomenon can occur for instance in case of acute myeloid leukemia described in the next section. It seems more realistic for stem cells to be fixed in a niche and the more a cell is mature, the more its ability to get detached from the colony would be important. This has not yet been taken into account in the software, but it is under current investigation. For the moment, all the cells have the same ability to move out from the bone marrow walls.

Multi-Agent Systems and Blood Cell Formation                                              405

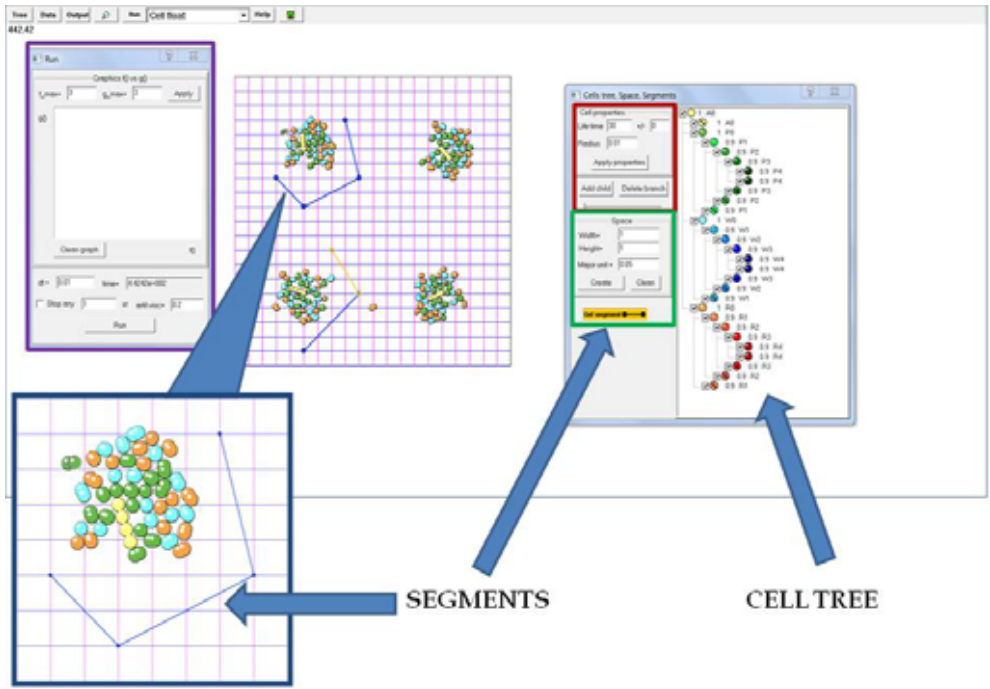

Fig. 4. Presentation of the different windows given by the software. The top left window represent the running application with different parameters such as viscosity. The left window gives the illustration of the tree consisting of all the cells and offspring, each differentiation probability are set up in this tree, each colour represents a cell type, the red frame corresponds to some of cell properties such as the life time, size (radius), etc. The green frame deals with the space property (size of the domain, and addition of segments). The small disks in the grids represent the different cells dividing, differentiating and dying with a focus on one part of the simulation at the bottom left of the figure.

### 4.2 Normal and pathological hematopoiesis

After setting up all the adjustments for the specific problem chosen by the user: number of cell lineages, cell fate, different probabilities (differentiation rates, apoptosis rates, etc.), size, bone marrow structure... It is possible to simulate normal and pathological hematopoiesis.

### 4.2.1 Normal hematopoiesis

To get an accurate model of normal hematopoiesis it is important to collect as many information about the different parameters as possible. This is the reason why it is necessary to exchange many discussions with hematologists. One attempt has been made, taking each size of cell type into account, with different proliferating times, apoptosis probabilities and different niches. The result has been plotted. However, many parameters need to be set up properly in good agreement with the experimental observations. This is under current investigation (see Fig. 5).

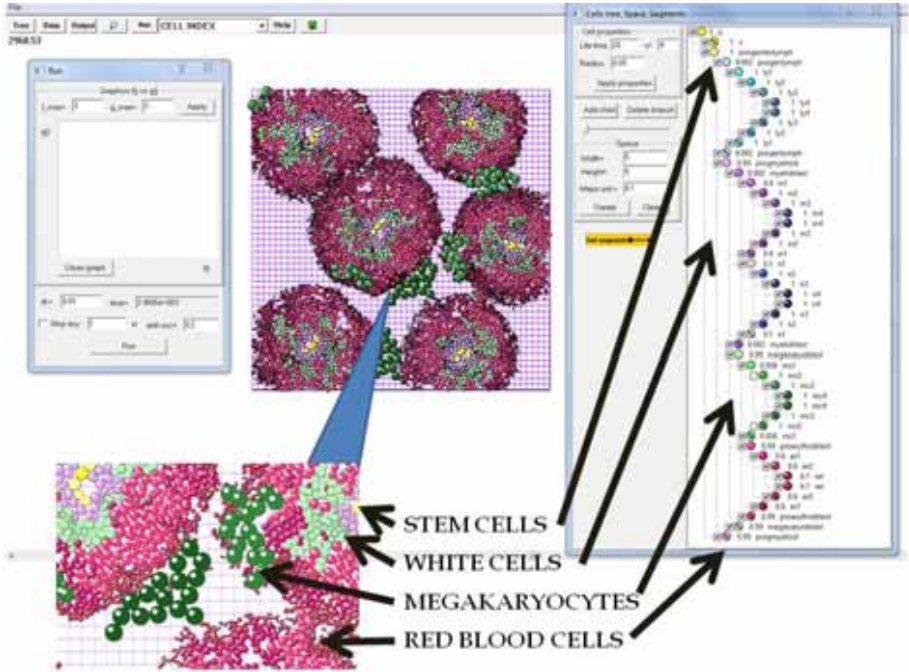

Fig. 5. Attempt of the simulation of a full normal hematopoiesis in one tiny part of the bone marrow. Each cell type has been determined with its size, colour, differentiation properties, etc. It is then possible to identify the dynamics for each lineage. One can observe the different colonies formed after the seeding of six hematopoietic stem cells.

**4.2.2 Pathological hematopoiesis: an application to leukemia**
Different pathologies could be simulated: stress erythropoiesis such as anemia, thrombocytopenia or leukemia amongst the most known possible diseases. We decide here to present only some results related to leukemia. We were interested in the different possibilities for leukemia to spread in the bone marrow. It is indeed known that some leukemia can be removed from the body without even the individual to be aware of suffering from the disease. However, in some cases, the pathology can settle down and spread until all the malignant cells fulfill the bone marrow. It can also be possible to get chronic symptoms of the disease. In other words, pathological cells and non pathological cell population would oscillate which could correspond to chronic myeloid leukemia cases.
Everything starts with a single cell that mutates. Settlement of the disease depends on different factors. Here are some of the main explanations of the pathological dynamics. The software allows the user to choose any cell to be a mutant cell. It can be a stem cell as well as any other cell. This cell can have the same properties as a normal cell but it can also have the ability to divide faster than the other, with a proliferation time much shorter than an average cell, or a division rate much more important. This cell can also differentiate and keep mutating to obtain more aggressive malignant cells. Here are three examples of the spread of the disease in the bone marrow.

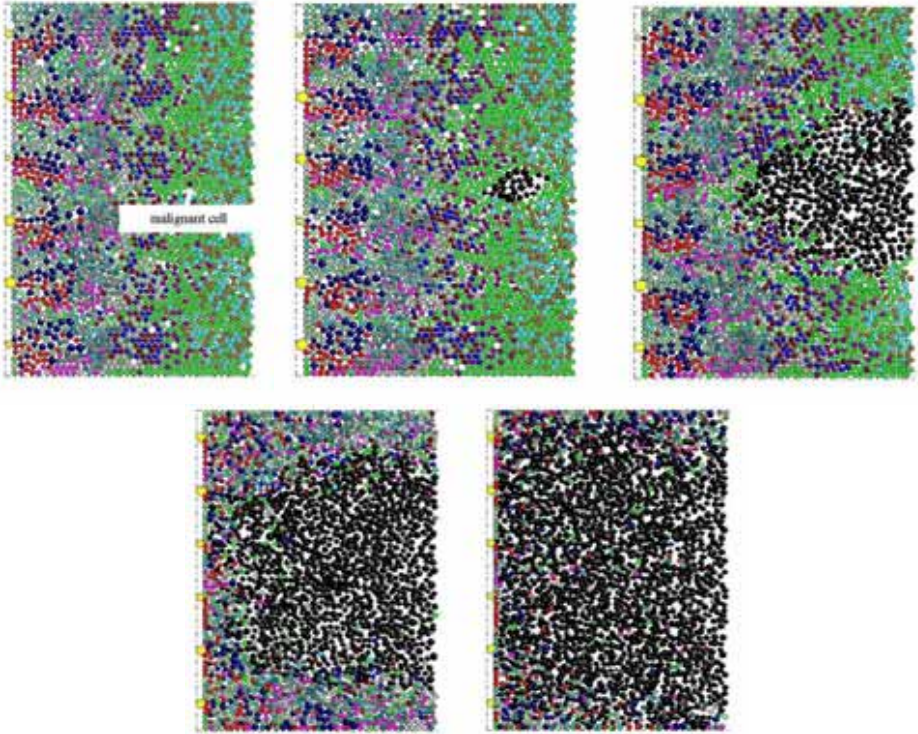

Fig. 6. Illustration of the spread of a disease. A malignant cell is placed in the bone marrow during normal hematopoiesis (top left) and develops to eventually invade the whole domain. This could correspond to a case of acute leukemia (illustration taken from Bessonov et al. 2006).

4.2.2.1 Acute leukemia

In this example, we consider a malignant cell that mutates. The cell chosen is not a stem cell. This choice was made in order to show that the disease can spread and settle down the marrow even if it is not a stem cell. It all depends on the properties given on the mutant cell. In this case, the pathological cells proliferate more rapidly than the average cells. At the early stage of the disease, it seems that leukemic cells will not stay in the marrow. But rapidly, they start to wash out the non pathological cells from the bone marrow and spread all around the place to eventually occupy almost all the medullar medium (see Fig. 6). It is important at this point to note that production of normal cells does not decrease in presence of the pathology. The only event occurring here is the strong local pressure that pushes other cells out to the blood systems. As a consequence, a large number of immature cells invade the blood stream which gives the onset of the symptoms. The proliferation time is of great importance in the spread velocity of the disease. It has been illustrated in Fig. 7. For a long proliferation time, malignant cells remain localized in a small region and do not seem to spread in the whole system. On the contrary, for a short proliferation time the disease would easily spread out and invade the bone marrow. A combination of different parameters would then allow the system to reach different equilibria depending on the threshold reached. A mathematical analysis corresponding to the simulation has not yet been done but should be investigated in the future showing a great range of dynamics. It is interesting for instance to note that if the density of stem cells is increased, and the same values for parameters are kept, then leukemia in these specific simulations has less chances to develop. Furthermore, if a mutation is given to a mature cell, this cell will have more difficulties to multiply and let the disease spread, and vice versa. Each choice could be driven by a specific type of leukemia one wants to model.

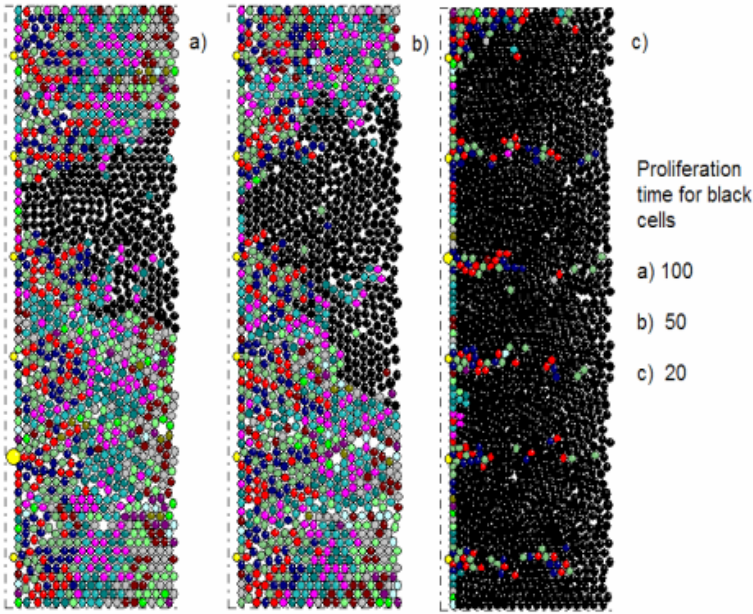

Fig. 7. Comparison between three simulations with different proliferation times for leukemic cells. The first one is modelled with a proliferation time of 100 time unit, the second 50, and the third 20 (illustration taken from Bessonov et al. 2006).

4.2.2.2 Chronic myelogenous leukemia (CML)

Chronic myelogenous leukemia occurs when concentration of cells of all types oscillate periodically. This kind of behavior has been observed in the 1970's and studied in the late 1990's and beginning of the 2000's (see section 3.1.3). It is possible, using a specific choice of parameters to obtain such oscillations with the Bessonov's software. Moreover, this application is able to get an output of the cell population leaving the bone marrow. Each cell type can be counted and put in a specific file. This file can be analyzed by any mathematical software able to analyze sets of data. The result we obtain with the example shown here seems qualitatively equivalent to the clinical data obtained in the 1970's (see Fig. 8). Getting quantitative results would be quite interesting for the biologist community in the sense that it would be then possible to replicate quite accurately the experimental or clinical data. This is still under investigation.

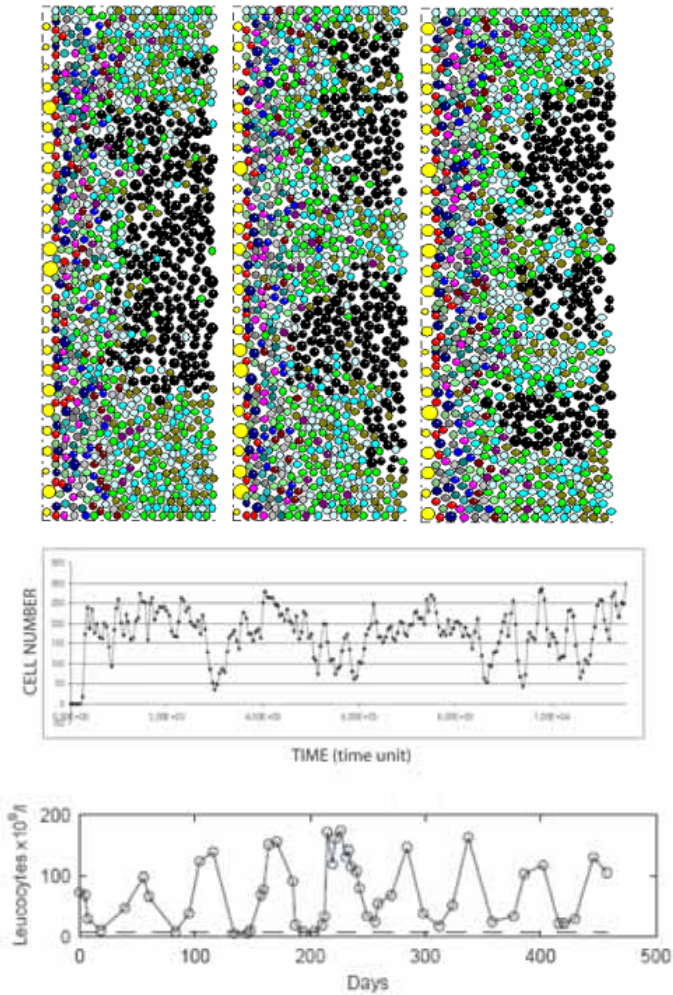

Fig. 8. Simulation of a pathological case. The black cells are malignant cells. The cell distributions are taken at three different times. The black cells form one, two or three domains almost periodically as shown in the time series right below the figure. It is possible to show that the concentrations of all cell types oscillate in time too (not shown here). The time series of the malignant cell population in the bone marrow could correspond to a CML case as illustrated by a time series of a leukocyte population in the bone marrow taken from a patient having CML (Fortin et al. 1999) (illustration taken from Bessonov et al. 2006).

4.2.2.3 A case where the disease fades out

Let us consider a case where the disease seems to spread in the bone marrow but eventually is being washed out and the systems returns to normal. In this example, the first generations of mutant cells can self-renew, and also differentiate to more aggressive cells. The last generations of mutant cells cannot self-renew. In this case, a simple simulation can exhibit a rapid increase of the pathology in the bone marrow. But after a certain period of time, the malignant cell population is washed out from the bone marrow and is replaced by normal cells (see Fig. 9). A question may arise then from this point: how is it possible for a population of mutant cells able to develop rapidly to disappear from the space representing the bone marrow here? The answer could enlightened by a simple focus of the individual level. In the simulation proposed here, all the stem cells are attached to the left wall of the domain, which is not the case for all the other cells. Thus, each cell except the stem cells is either condemned to die by apoptosis, differentiate, self-renew or leave the blood stream after a certain period of time. This is also the case for the cells defined as mutant cells in the example here. They are also part of the dynamics rules of the software. They can self-renew or differentiate but do not increase their number. It may thus be impossible for these cells to overcome their loss and eventually, they are pushed out of the bone marrow by the younger generations of healthy cells. In some cases however, self-renewing mutant cells can divide with a rate large enough to spread out and settle down in the medulla.

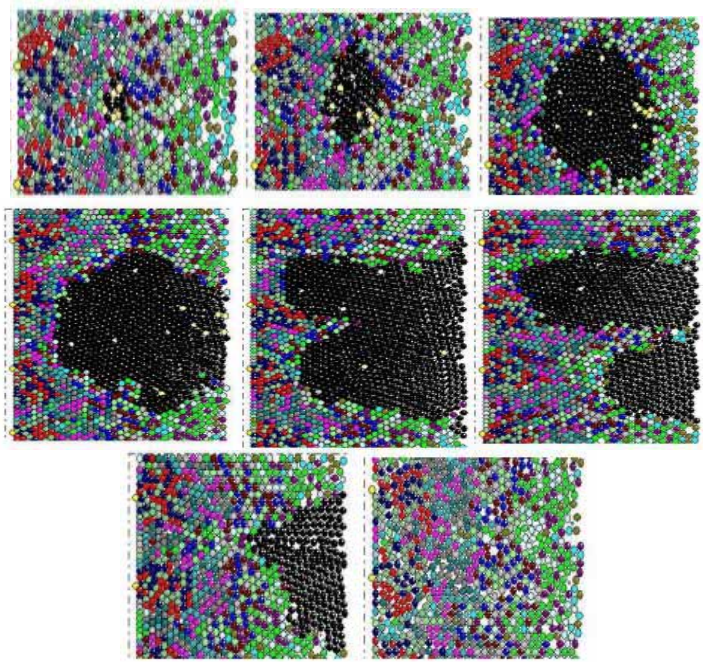

Fig. 9. Evolution of the disease spread: malignant (black) cell population grows from a malignant focus but not for a long time but eventually get washed out by the healthy cells. The disease free steady state is then stable in this example. This could correspond to what happens frequently in everyone's body (illustration taken from Bessonov et al. 2006).

In the next section, we go further into the cell environment, by taking the cell communication into account. Thus, not only space competition in the bone marrow plays a role for the development of different cells, but also the influence of the environment and the communication between cells in one neighborhood via some exchange of molecules.

**4.3 Cell communication**

Cell communication corresponds to another application of the Bessonov's software. In this section, we give an overview of the possibilities given by this application and the influence of different parameters on the cell population dynamics.

**4.3.1 How does cell communication is taken into account in the software?**

4.3.1.1 A simple example

It is known now that cell differentiation, self-renewal and apoptosis properties can be ruled by complex dynamics of molecules produced inside and outside each cells. For instance, it has been discovered that the stimulation hormones like EPO (see 1.5) would decrease the apoptosis rate in the red blood cell lineage. On the other hand, cell differentiation and the choice of one of the lineages can be regulated by a system of transcription factors. Some mathematical models have been attempted to tackle this problem (Roeder 2006, Crauste 2010). In Huang et al. (2007), the authors developed a model of binary cell fate decisions combining stochastic and deterministic instructions. In our work, we decided to give the possibility for the user to add this ability of cell differentiation through communication with the environment to the existing other applications provided by our software and described above. As mentioned by Roeder et al. (2006a) lineage specification is "a competition process between different interacting lineages propensities".

Even if our software allows the user to simulate several cell lineage specifications and communication with as many stimulating factors as wanted, we believe that a description of the application use with the simplest model of only two subpopulations would be more understandable.

4.3.1.2 An exchange of information

Let us then start with one population of undifferentiated cells denoted by A. These cells can divide, giving two daughter cells. One of the daughter cell would be exactly of the same type of its mother (self-renewing), and the other would be of either type G or type F. Two possible lineages are then given to the undifferentiated cells. The color given to the A-type cell would be white, if would be blue for the G-type and red for the F-type (see Fig. 10). Each cell denoted i- not even a cell type but really each individual - is characterized by two functions: fi and gi depending on time the time t, which could correspond for instance to a certain amount of two types of molecules. We assume in this simulation that every new born cell is undifferentiated, that is white. In other word, every time that a stem cell divides, it gives rise to two white cells. These cells are prescribed by the same initial amount of molecules, *i.e.* $f_0$, $g_0$. This content changes with time with a rate defined by two differential equation describing the evolution of $f_i$ and $g_i$ with respect to time:

$$\frac{df_i}{dt}=a(F_i-f_i), \text{ and } \frac{dg_i}{dt}=a(G_i-g_i), \qquad (1)$$

where a is a constant, and the $F_i$ and $G_i$ can be chosen to satisfy the so-called "average rule", that is

$$F_i=\sum_{j\neq i} f_j/N, \text{ and } G_i=\sum_{j\neq i} g_j/N, \qquad (2)$$

where N is the sum of the number N of neighbor cells, which is defined automatically in the software code. The other possibility to define Fi and Gi is to follow the "max rule", that is the cell content fi will evolve depending on the influence of its neighbor having the greatest amount of F-type molecule and likewise for gi. In other words,

$$Fi = \max_{j \neq i} fi \text{ and } Gi = \max_{j \neq i} gi. \tag{3}$$

From a biological point of view, this means that each individual releases the molecules of F or G-type with a rate proportional to their concentration, and similarly, it receives the molecules from its neighboring cells with a rate proportional to their concentrations.

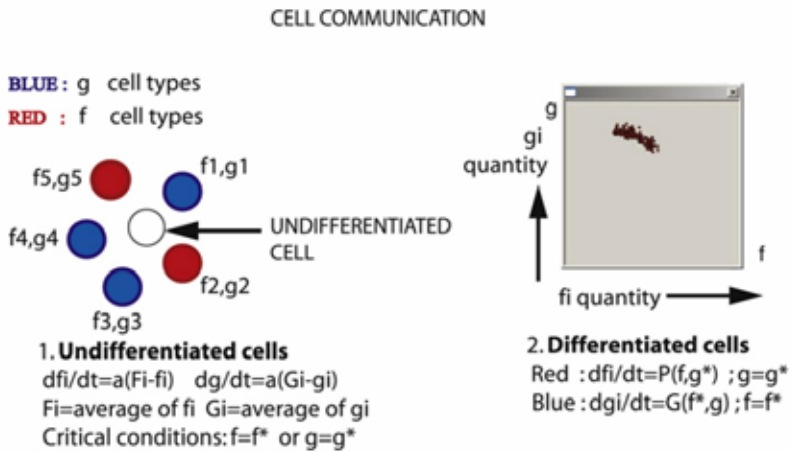

Fig. 10. Cell communication: each cell denoted "i" at a time t is characterized by two parameters, fi(t) and gi(t). Under specific conditions the neighboring cells can influence the undifferentiated (white) ones. This influence depends on the number of each cell type around. After a certain time, the simulation is stopped, the software counts the number of each cell type (f and g) and plot their distribution in the f − g plane (illustration taken from Bessonov et al. 2006).

4.3.1.3 Choice of one lineage and evolution of the cell content

Different rules are set up for an undifferentiated cell to remain white or to choose a lineage. A new born cell stays white as long as $fi^2(t) + gi^2(t) \leq \sigma$, where $\sigma$ is a given parameter set up by the user. This means that a cell needs a specific threshold of molecule concentration to be differentiated taking the color of the greater concentration between fi and gi to become red or blue.

When a cell has "chosen" its lineage, its content can evolve depending on the following rules:

$$\frac{dfi}{dt} = P(fi, gi), \ gi = gi^* = \text{constant} \tag{4}$$

for red cells, and

$$\frac{dgi}{dt}=Q(fi,gi), \ fi=fi^*=\text{constant} \tag{5}$$

for blue cells, where fi* and gi* stand for the concentration of molecules of F and G-type at the moment when differentiation has been decided. In other words, if a cell is of type F, the amount of "F-molecules" would change depending on the defined function P but the "G-molecule" content will remain constant and vice versa with function Q. This property gives the specific shape of the figures representing the simulations on the cell type evolution with a blank squared shape on the upper right part of the plot (see Fig. 11). The function P and Q are defined to be quadratic functions as follows

$$P(f,g)=a_1+a_2f+a_3f^2+a_4fg+a_5g, \tag{6}$$

and

$$Q(f,g)=b_1+b_2g+b_3g^2+b_4fg+a_5f, \tag{7}$$

where ai and bi, i=1,...,5 are some constants defined by the user. The quadratic form of P and Q was chosen for simplicity, but can be modified anytime by the user.

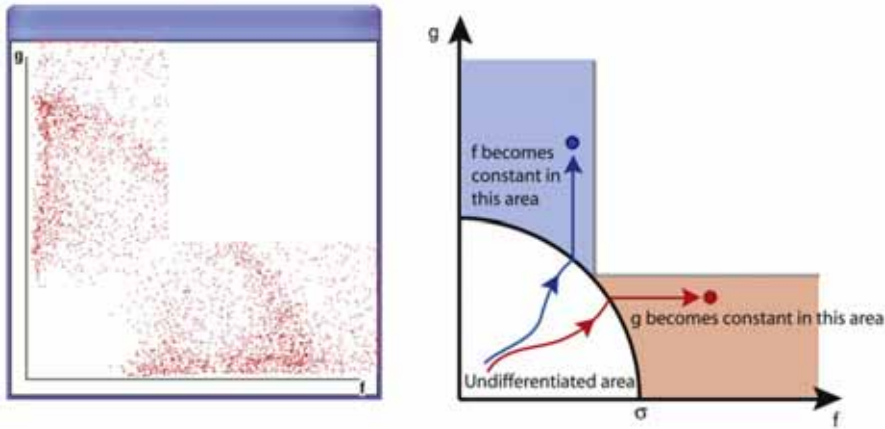

Fig. 11. Cell position in the f–g plane. A white cell remains undifferentiated as long as $fi^2(t)+gi^2(t) \leq \sigma$ (inside the disk (figure in the left)), where σ is a given parameter. If the concentrations of f and g become sufficiently high (greater than σ), the cell chooses its type. Once the differentiation occurs and the cell chooses its type, further evolution of f and g becomes different. For red cells df i/dt =P(fi,gi), gi = g*i (it remains constant), for blue cells dgi/dt = Q(fi,gi), fi=f*i (it remains constant). In other words, after differentiation, the value of f in red cells increases and the value of g remains constant; for blue cells g increases and f remains constant which gives the specific shape of the f − g graph (right) (illustration taken from Bessonov et al. 2009).

In each figure it is then possible to represent the cell of all types with undifferentiated cells on the quarter of disk on the bottom left angle with radius equal to σ (corresponding to the distance between the origin of the axis and the bottom left part of the blank squared shape.

Everything above this disk, between the y-axis and the blank squared shape would correspond of cells of blue G-type and the higher the cell is located the more mature it is. On the other hand, all cells on the right side after the disk and below the blank squared shape are the red F-type cells. And thus, the choice of the lineage can be well defined depending on the zone a cell can be plot. Consequently, if all cells remain within the disk, this means that all cells remain undifferentiated; this could correspond to a case of acute leukemia. They can also be located in one or two lineages, or the three depending on the sets of parameters chosen. This will be developed in the next paragraph.

Furthermore, cell generations can be observed by the "circular stripes" appearing in the simulations. This is correlated to the ratio of the proliferation time between the stem cells and the first daughters. Let us see some examples in the next paragraph.

### 4.3.2 Examples of cell communication and differentiation

We remind the reader here the starting bases of the cell communication application. At the beginning of the simulation, only undifferentiated cells are produced by the stem cells. Once the whole domain has been filled up with all the white cells, the process stops and each cell is prescribed randomly one of the two (red or blue) types with some value fi and gi. The application starts again and then, all new born undifferentiated cells are obliged to choose one of the types depending on their environment and the parameters set up as explained in the previous paragraph. Some structures can appear. This process starts with a random distribution of the cells, but specific structures can appear depending on the different sets of parameters.

The application can give different outputs: the number of cells of each type can appear in a specific file as explain in a previous section. In other words, after a certain time $\tau_i$, which represents the moment when the ith cell leaves the bone marrow. The cell is then registered into the file with its fi and gi content, which determines its type depending on the level of red or blue molecule inside. To be more convenient, the software plots directly all these cells on the (f,g)-plane. In other words, it $(f(\tau_i),g(\tau_i))$ plots corresponding to each cell leaving the bone marrow. The graph obtains represent then the population of blood cells released in the system. Let us give three examples corresponding to the influence of the main parameters: starting with the proliferation time, then the cell size, and finally a combination of the cell communication parameters with the self-renewal processes.

4.3.2.1 Cell communication and the proliferation time parameter

In this example, let us assume that cell of the second generation cannot differentiate anymore. This may represent a simplified case of normal hematopoiesis. Almost no undifferentiated cells are found in the bone marrow, and a great proportion of cells clearly belong to one of the two types and increasing the proliferation time would improve the cell differentiation process. It is possible to give a biological explanation behind these simulations. It is indeed, easy to understand that in the case of non pathological hematopoiesis, a cell being given more time to mature will leave the bone marrow with a more complete functional material than cells having little time to fulfill the maturation process. It has been shown for instance that in case of stress erythropoiesis (like anemia or blood loss), cell proliferation is accelerated due to the combination of self-renewal process of progenitors, but also the apoptosis rate decreases in the bone marrow, due to the effect of EPO, but once in the blood stream, cell death is greater than normal due to the fact that they have not completed their formation and are released in the stream with a weaker structure. Moreover, changing the distribution of time (the +/- time column) would imply a change in the sharpness of cell generation plots in the sense that the zones representing each generation does not show clear borders when the distribution of time is increased ((see Fig. 12). This makes sense since the cell proliferation occurs more randomly in time. The process tends to be then more stochastic.

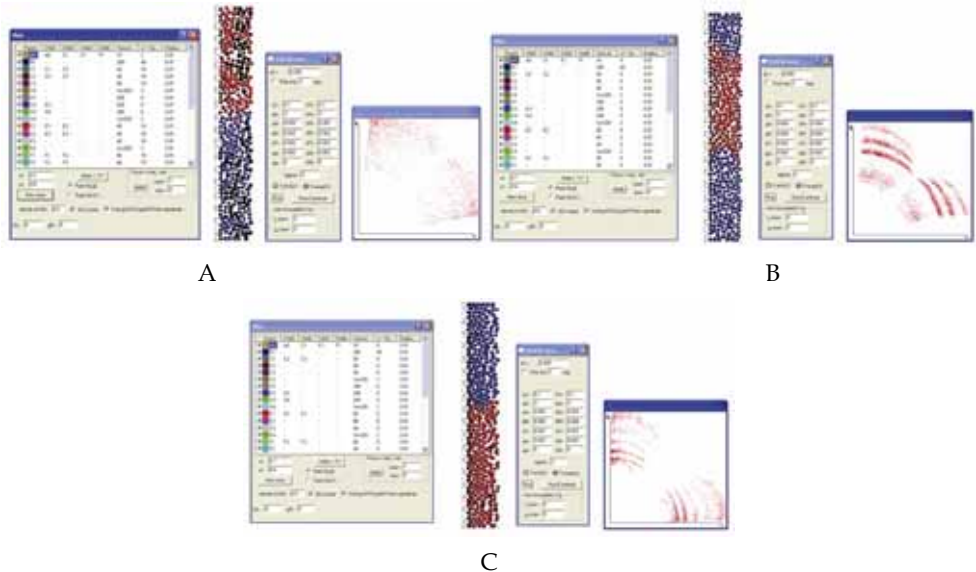

Fig. 12. Changes in cell proliferation time. A. In this simulation, the proliferation time is not exact but distributed. Thus in the plot on the f –g plane one can observe a less structured differentiated cell distribution. B. In this simulation, the cell lineages have been modified: no offspring for several cell generations. In the plot on the f –g plane one can observe then a well defined structure of all the cell generations depending on their maturity level as well as their differentiation profile. This could correspond to a non pathological case. C. Changes in cell proliferation time that gives a simulation where almost no cells are undifferentiated and cell generations are well separated (illustration taken from Bessonov et al. 2009).

4.3.2.2 Cell communication and the cell size parameter

As explained in the paragraph 4.1.1.1., the cell size decreases in correlation with their maturity except for the megakaryocytic lineage. Thus, it appears necessary to consider the influence of the cell size parameter in our application. When comparing the simulations with the size decrease taken into account, one can see that no undifferentiated cells can be found in the blood stream. All of them remain in the bone marrow (in the non pathological case considered here). On the other hand, the cell generations are clearly defined. This result was expected since, with increasing maturity, the cells get smaller, they need less space in the bone marrow, and so they can stay longer than the one whose size has not changed, and thus can gain more differentiation molecules (red or blue). And thus, the f and g colonies appear much more distinct than a simple simulation where no evolution of size is taken into account (see Fig. 13).

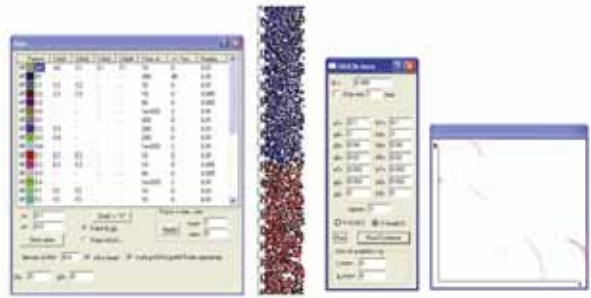

Fig. 13. Changes in cell radius in order to get smaller size as maturity increases, gives clear and definite compartments of cell generations (illustration taken from Bessonov et al. 2009).

4.3.2.3 Cell communication and the self-renewal process

Let us assume that the system is under a stress erythropoiesis such as a severe anemia or blood loss. Then it has been shown that progenitor cells and sometimes precursor cells can self-renew under stimulating glucocorticoids. What is expected with this change in the parameter set up is an increase of undifferentiated. The population is indeed forced to increase and to give rapidly efficient cells to be released in the blood stream. This process could also be pushed to and extreme case, where most of the cells leaving the bone marrow could be undifferentiated. This would then correspond to cases of acute leukemia, where it is not possible for cells to differentiate, and thus malignant cells would only self-renew with almost no differentiation. The system, after a certain time would then be filled with undifferentiated cells only (see Fig. 14).

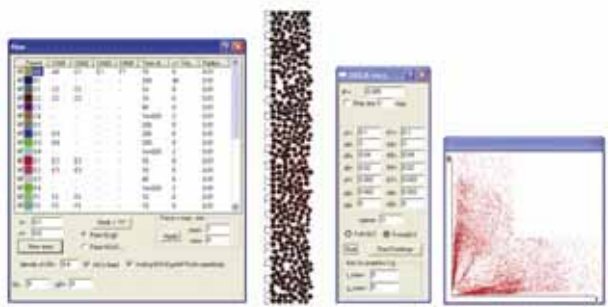

Fig. 14. Adding self-renewal ability in progenitors leads to a nice illustration of what could occur in case of severe bleeding or anemia, and action of the glucocorticoids on cell differentiation (illustration taken from Bessonov et al. 2009).

### 4.4 Niches and colonies in the bone marrow

In this section, the mechanical interaction and the cell displacement are the same as in the previous software interface. However, instead of imposing self-renewal, differentiation and apoptosis like it is in the previous model (even if it is a stochastic process in the software, it is set up by the user through different parameter choices), these properties are ruled out by intrinsic and extrinsic regulation with no user implication.

### 4.4.1 Hybrid models

Hereafter we propose a new approach of mechanical interactions between cells both from intrinsic and extrinsic regions. In order to do so, we focus our attention on intra-cellular regulation, described by ordinary differential equations, and on extra-cellular regulation, described by partial differential equations.

We restrict ourselves here to the simplest model where cells are represented as elastic balls. In other words, we consider two elastic balls with their centres at the points x1 and x2, and with the radii, respectively given by r1 and r2. If the distance d12 between the centres is less than the sum of the radii, r1 + r2, then there is a repulsive force between them, denoted by f12 depending on the distance d12. Moreover, if a particle with the centre at xi is surrounded by several other particles with their centres at the points xj, j=1,..., k, then we consider the pair wise forces fij assuming that they are independent of each other. This assumption corresponds to small deformation of the particles. Hence, we find the total force Fi acting on the i-th particle from all other particles, $F_i = \sum_{j \neq i} f_{ij}$. The motion of the particles can now be described as the motion of their centers. By Newton's second law

$$m\ddot{x}_i + \mu m \dot{x}_i - \sum_{j \neq i} f(d_{ij}) = 0, \qquad (8)$$

where m is the mass of the particle, the second term in the left-hand side describes the friction by the surrounding medium. Dissipative forces can also be written in a different form. This is related to dissipative particle dynamics (Karttumen, 2004). Intra-cellular regulatory networks for the i-th cell are described by a system of ordinary differential equations

$$\frac{du_i}{dt} = F(u_i, u_e), \qquad (9)$$

where ui is a vector of intra-cellular concentrations, ue is a vector of extra-cellular concentrations, F is the vector of reaction rates which should be specified for each particular application. Evolution of the concentrations of the species in the extra-cellular matrix is described by the diffusion equation

$$\frac{\partial u_e}{dt} = D\Delta u_e + G(u_e, c), \qquad (10)$$

where c is the local cell density, G is the rate of consumption or production of these substances by cells. These species can be either nutrients coming from outside and consumed by cells or some other bio-chemical products consumed or produced by cells. In particular, these can be hormones or other signalling molecules that can influence intra-cellular regulatory networks. In some cases, convective motion of the medium should be taken into account.

### 4.4.2 1-D model example

We begin with the 1D model example where cells can move along the line. The coordinates x i in equation (1.1) are real numbers. Each cell can divide or die by apoptosis. After division a cell gives birth to two daughter cells identical to itself (this is the case of self-renewal division, there is no differentiation taken into account in this example). We suppose that cell division and death are influenced by some bio-chemical substances produced by the cells themselves.

We consider the case where there are two such substances, whose concentrations are denoted by ue and ve and satisfy the following system of equations:

$$\begin{cases} \dfrac{du_e}{dt} = d_1 \dfrac{\partial^2 u_e}{\partial x^2} + b_1 c - q_1 u_e, \\ \dfrac{dv_e}{dt} = d_2 \dfrac{\partial^2 v_e}{\partial x^2} + b_2 c - q_2 v_e. \end{cases} \quad (11)$$

These equations describe the evolution of the extracellular concentrations ue and ve with their diffusion, production terms proportional to the concentration of cells c and with the degradation terms. We note that cells are considered here as point sources with a given rate of production of u and v. The cell concentration is understood as a number of such sources in a unit volume. In numerical simulations, where cells have a finite size, we consider them as distributed sources and specify the production rate for each node of the numerical mesh. Intra-cellular concentrations ui and vi in the i-th cell are described by the equations:

$$\begin{cases} \dfrac{du_i}{dt} = k_1^{(1)} u_e(x,t) - k_2^{(1)} u_i(t) + H_1, \\ \dfrac{dv_i}{dt} = k_1^{(2)} v_e(x,t) - k_2^{(2)} v_i(t) + H_2. \end{cases} \quad (12)$$

Here and in what follows we write equations for intra-cellular concentrations neglecting the change of the cell volume. This approximation is justified since the volume changes only twice before cell division and this change is relatively slow. The first term in the right-hand side of the first equation shows that the intra-cellular concentration ui grows proportionally to the value of the extra-cellular concentration ue (x, t) at the space point x where the cell is located. It is similar for the second equation. These equations contain degradation terms and constant production terms, H 1 and H 2. When a new cell appears, concentrations ui and vi are set to zero.

If the concentration ui reaches some critical value uc, then the cell divides. If vi reaches vc, the cell dies. Consider first the case where

$$k_1^{(1)} = k_1^{(2)} = k_2^{(1)} = k_2^{(2)} = 0. \quad (13)$$

Then ui and vi are linear functions of time which reach their critical values at some times t=τu and t = τv,

respectively. If τ u < τ v , then all cells will divide with a given frequency, if the inequality is opposite, then all cells will die.

Next, consider the case where $k_1^{(1)}$ is different from zero (see Bessonov, 2010 for other examples). If it is positive, then cells stimulate proliferation of the surrounding cells, if it is negative, they suppress it. Both cases can be observed experimentally. We restrict ourselves here by the example of negative $k_1^{(1)}$. All other coefficients remain equal to zero. Therefore, cells have a fixed life time τ v . If they do not divide during this time, they die.

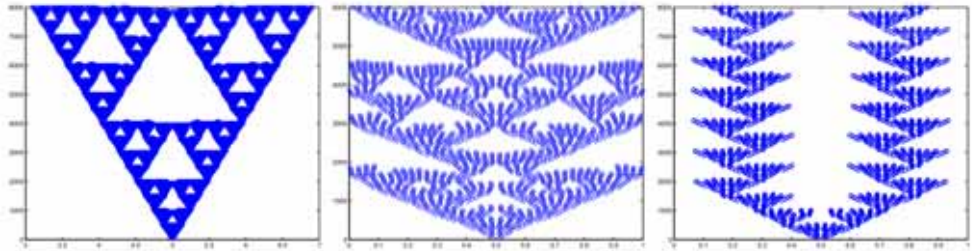

Fig. 15. Dynamics of cell population in the case where cells either self-renew or die by apoptosis. Cells are shown with blue dots. Horizontal axis shows cell position, vertical axis shows time.

We carry out the 1D simulation where cells can move along the straight line. Initially, there are two cells in the middle of the interval. Figure 15 shows the evolution of this population in time. For each moment of time (vertical axis) we have the positions of cells (horizontal axis) indicated with blue points.

The evolution of the cell population in Figure 15 (left) can be characterized by two main properties. First of all, it expands to the left and to the right with approximately constant speed. Second, the total population consists of relatively small sub-populations. Each of them starts from a small number of cells. Usually, these are two cells at the right and at the left of the previous sub-population. During some time, the sub-population grows, reaches certain size and disappears giving birth to new sub-populations.

This behavior can be explained as follows. The characteristic time of cell division is less than the one of cell death. When the sub-population is small, the quantity of $u_e$ is also small, and its influence on cell division is not significant. When the sub-population becomes larger, cell division is slowed down because of growth of $u_e$. As a result the sub-population disappears. The outer cells can survive because the level of $u_e$ there is less.

The geometrical pattern of cell distribution for these values of parameters reminds Serpinsky carpet (Figure 15, left), an example of fractal sets. The pattern of cell distribution depends on the parameters. Other examples are shown in Figure 15 (middle and right).

The simulations presented here do not use the extra-cellular variable $v_e$. Instead of the variable $u_e$, which decelerates cell proliferation, we can consider $v_e$ assuming that it accelerates cell apoptosis. In this case, qualitative behavior of cell population is similar.

### 4.4.3 Erythropoiesis modeling

We consider two types of erythroid cells, progenitors and reticulocytes. Erythroid progenitor fate (differentiation, self-renewal, death by apoptosis) is supposed to be regulated by intra-cellular mechanisms (protein competition) and extra-cellular substances. The main external source of control in what follows will be Fas-ligand, a membrane protein produced by reticulocytes that activates the intra-cellular protein Fas. Other extra-cellular substances include EPO and glucocorticoids, among others. We will restrict our model to the influence of EPO, and even though we do not detail this action in the following, the level of EPO can be considered either constant or proportional to mature erythrocyte quantity. Fas-ligand will act on progenitor differentiation and apoptosis, whereas EPO will inhibit progenitor apoptosis and increase self-renewal.

We assume intracellular regulation of erythroid progenitors is determined by two proteins, ERK and Fas (Crauste, 2010), although several other proteins may play a role in this regulation. A simplified model is given by the system of two ordinary differential equations (Crauste, 2010)

$$\frac{dE}{dt} = \left(\alpha(EPO) + \beta E^k\right)(1-E) - aE - bEF, \tag{14}$$

$$\frac{dF}{dt} = \gamma(F_L)(1-F) - cEF - dF, \tag{15}$$

where E and F are intra-cellular concentrations of ERK and Fas, a, b, c, d are some non-negative parameters, $\alpha$ is a function of erythropoietin (EPO) and $\gamma$ is a function of Fas-ligand, whose concentration is denoted by $F_L$. For fixed values of EPO and $F_L$, (14)-(15) is a closed system of ordinary differential equations. It can have from one to three stationary points. Its detailed analysis is presented in (Crauste, 2010).

The concentration of Fas-ligand is described by the diffusion equation

$$\frac{dF_L}{dt} = d\Delta F_L + W - \sigma F_L, \tag{16}$$

where W is a source term proportional to the concentration of reticulocytes. Though Fas-ligand is considered as a surface protein, and its interaction with erythroid progenitor basically occurs when they are in physical contact with reticulocytes, we model it as if it could diffuse in the extracellular matrix. If the diffusion coeffcient is suffciently small, it is located in a small vicinity of reticulocytes. Therefore, Fas-ligand influences erythroid progenitors when they are suffciently close to reticulocytes.

Let us summarize the model. System (14)-(15) is considered inside each erythroid progenitor with its proper initial condition (see below) and with the value of $F_L$ which can depend on its spatial location and on time. Erythroid progenitors can proliferate or die by apoptosis. Apoptosis occurs if the intracellular Fas concentration reaches some critical value $F_c$. In this case, the cell is removed from the computational domain.

If the cell does not die by apoptosis, then it proliferates, that is it divides at the end of cell cycle. Cell cycle is composed of two parts: G0/G1 phases and S/G2/M phases. The duration of the G0/G1 phase is chosen randomly from 0 to some maximal value $\tau_{max}$ with the typical values $6 - 12$ hours, the duration of the remaining part of cell cycle is fixed, usually 12 hours.

Cell proliferation can result in self-renewal or differentiation. In the first case, the two daughter cells are also erythroid progenitors. For each of them we consider intracellular regulation with system (14)-(15). The values of ERK and Fas in the newly born cells can either be some given parameters or equal half those of the mother cell. In the case of differentiation, the two daughter cells become reticulocytes. The choice between self-renewal and differentiation is determined by the values of ERK in the process of cell cycle. Once it reaches a critical value $E_c$, the cell self-renews. Otherwise, it differentiates. These assumptions are in agreement with actual biological understanding of these processes. We do not consider intracellular regulation for reticulocytes. Once they appeared, they remain in the computational domain one cell cycle more in order to become mature erythrocytes. Then they are removed. This corresponds to the fact that erythrocytes leave the bone marrow to enter blood flow. Reticulocytes produce Fas-ligand with a constant rate. Fas-ligand influences intracellular regulation of erythroid progenitors through equation (15). It increases Fas production rate resulting in apoptosis of the progenitors if Fas concentration is sufficiently high or in their differentiation for intermediate values of F L . For greater values of Fas-ligand, trajectories of system (14)-(15) move towards greater values of F and to smaller values of E. Hence, the critical value of ERK may not be reached and the cell will differentiate.

Let us finally recall that proliferation and apoptosis change cell spatial distribution. New cells, when they appear, push each other creating cell displacement (see Fig. 16).

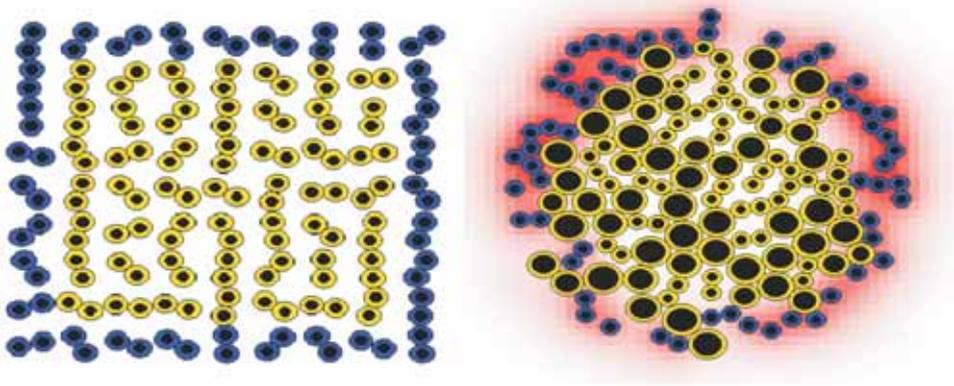

Fig. 16. Modelling of erythroblastic islands: initial cell distribution (left) and a stable island (right). Yellow cells in the center are erythroid progenitors, blue cells at the border are reticulocytes producing Fas-ligand (in red).

## 5. Conclusion

In this chapter, our goal was to give an insight of the different attempts to model the blood cell formation: several purely deterministic, some purely stochastic, few taking the space structure of the bone marrow and thus space competition into account. It appeared important to us to develop a compromising model where stochasticity is mixed with the medulla structure. This approach using the multi-agent systems seems to us a good way to describe different mechanism related to the normal and pathological hematopoiesis. We have shown that with our software it was possible to get a rich variety of behaviors. Our simulations found some interest in the community of biologists and clinicians. However, even if we obtained some relevant results none of them where quantitatively compared with experimental data. This is part of the perspective work taken in consideration. Another objective would be to consider the stimulation factors more explicitly by closing with the use of different feedbacks.